\documentclass[%
aip,
preprint,
showpacs,
preprintnumbers,
 amsmath,
 amssymb,
]{revtex4-1}

\usepackage{graphicx}
\usepackage{dcolumn}
\usepackage{bm}
\usepackage{color}

\begin{document}

\title{Growth, saturation and collapse of laser-driven plasma density gratings}

\author{H. H. Ma}
\affiliation{Key Laboratory for Laser Plasmas (MoE), School of Physics and Astronomy, Shanghai Jiao Tong University, Shanghai 200240, China}%
\affiliation{Collaborative Innovation Center of IFSA, Shanghai Jiao Tong University, Shanghai 200240, China}%

\author{S. M. Weng}\email{wengsuming@gmail.com}%
\affiliation{Key Laboratory for Laser Plasmas (MoE), School of Physics and Astronomy, Shanghai Jiao Tong University, Shanghai 200240, China}%
\affiliation{Collaborative Innovation Center of IFSA, Shanghai Jiao Tong University, Shanghai 200240, China}%

\author{P. Li}
\affiliation{Research Center of Laser Fusion of China Academy of Engineering Physics, Mianyang, SiChuan, 621900, China}%

\author{X. F. Li}
\affiliation{Key Laboratory for Laser Plasmas (MoE), School of Physics and Astronomy, Shanghai Jiao Tong University, Shanghai 200240, China}%
\affiliation{Collaborative Innovation Center of IFSA, Shanghai Jiao Tong University, Shanghai 200240, China}%

\author{Y. X. Wang}
\affiliation{Key Laboratory for Laser Plasmas (MoE), School of Physics and Astronomy, Shanghai Jiao Tong University, Shanghai 200240, China}%
\affiliation{Collaborative Innovation Center of IFSA, Shanghai Jiao Tong University, Shanghai 200240, China}%

\author{S. H. Yew}
\affiliation{Key Laboratory for Laser Plasmas (MoE), School of Physics and Astronomy, Shanghai Jiao Tong University, Shanghai 200240, China}%
\affiliation{Collaborative Innovation Center of IFSA, Shanghai Jiao Tong University, Shanghai 200240, China}%

\author{M. Chen}
\affiliation{Key Laboratory for Laser Plasmas (MoE), School of Physics and Astronomy, Shanghai Jiao Tong University, Shanghai 200240, China}%
\affiliation{Collaborative Innovation Center of IFSA, Shanghai Jiao Tong University, Shanghai 200240, China}%

\author{P. McKenna}
\affiliation{SUPA, Department of Physics, University of Strathclyde, Glasgow G4 0NG, UK}%

\author{Z. M. Sheng}
\affiliation{Key Laboratory for Laser Plasmas (MoE), School of Physics and Astronomy, Shanghai Jiao Tong University, Shanghai 200240, China}%
\affiliation{Collaborative Innovation Center of IFSA, Shanghai Jiao Tong University, Shanghai 200240, China}%
\affiliation{SUPA, Department of Physics, University of Strathclyde, Glasgow G4 0NG, UK}%
\affiliation{Tsung-Dao Lee Institute, Shanghai 200240, China}%
\date{\today}
\begin{abstract}
The plasma density grating induced by intersecting intense laser pulses can be utilized as optical compressors, polarizers, waveplates and photonic crystals for the manipulation of ultra-high-power laser pulses. 
However, the formation and evolution of plasma density grating are still not fully  understood as linear models are adopted to describe them usually.
In this paper, two theoretical models are presented to study the formation process of plasma density grating in the nonlinear stages.
In the first model an implicit analytical solution based on the fluid equations is presented, while in the second model a particle-mesh method is adopted.
It is found that both models can describe the plasma density grating formation at different stages, well beyond the linear growth stage.
More importantly, the second model can reproduce the phenomenon of ion ``wave-breaking" of plasma density grating, which eventually induces the saturation and collapse of plasma density grating.
Using the second model, the saturation time and maximum achievable peak density of plasma density grating are obtained as functions of laser intensity and plasma density, which can be applied to estimate the lifetime and capability of plasma density grating in experiments.
The results from these two newly-developed models are verified using particle-in-cell simulations.
\end{abstract}


\maketitle

\section{Introduction}

Since the invention of the chirped pulse amplification technology \cite{Strickland}, laser peak power and focused intensity have increased many orders of magnitude in the last three decades.
Nowadays, there are a number of laser systems in the world that can deliver petawatt (PW) laser pulses \cite{Danson}, which can be tightly focused to ultrahigh intensities $\sim 10^{21}$ W/cm$^2$.
The interactions of such intense laser pulses with materials bring about rich physical phenomena and many prospective applications \cite{MourouRMP,Gibbon,Weng_S}.
However, the manipulation of such laser pulses becomes more and more challenging for conventional solid-state optical components, which are susceptible to optical damage at high laser energy intensities.
For silica, which is the most widely used material in solid-state optics, the laser-induced damage threshold of energy fluence is on the level of 10 J/cm$^2$ in the femtosecond to picosecond regime.
In order to keep the laser energy fluence below this damage threshold, the diameters of solid-state optical components are usually required to be meter-scale for multi-PW laser systems.
In contrast, plasmas resulted from the ionization of materials can sustain much higher laser intensities than solid crystals.
Consequently, plasma-based optical components for the manipulation of ultra-high-power laser pulses can be made much more compact than their conventional solid-state optical components.
As a result, plasma-based optics are attracting growing attention \cite{Doumy,Thaury,Malkin,Trines,Weber,Lehmann2013,Michel2009,Marion,Weng2017,Zheng2019,Edwards,Sheng,WuAPL,WuPoP,Michel2014,Turnbull,Lehmann2016,Lehmann2018}.

To date, a lot of novel schemes based on plasma optics have been proposed for the manipulation or amplification of intense laser pulses.
Plasma mirrors are widely used for enhancing the temporal contrast of intense laser pulses \cite{Doumy,Thaury},
Raman or Brillouin scattering in laser-plasma interactions are studied for the amplification of laser pulses \cite{Malkin,Trines,Weber,Lehmann2013},
cross-beam energy transfer in plasmas is studied for tuning the implosion symmetry of inertial confinement fusion targets \cite{Michel2009,Marion},
and magnetized plasmas are proposed for the polarization control of ultra-high-power laser pulses \cite{Weng2017,Zheng2019} or the amplification of intense laser pulses \cite{Edwards}.
In particular, two intersecting intense laser pulses in a plasma can induce a plasma density modulation and form a periodic density structure, i.e., a plasma density grating (PDG) \cite{Sheng}.
Such a PDG can also be produced via ponderomotive steepening due to the interference between the incident and reflected laser pulses in laser-plasma interactions \cite{RSmith}.
The PDG can sustain a relatively high laser intensity and exist in a quasi-steady state for several picoseconds.
Therefore, it becomes an attractive approach for the manipulation of femtosecond intense laser pulses, and is studied for broad applications such as the plasma compressor, the plasma polarizer and waveplate, and the transient plasma photonic devices for high-power laser \cite{WuAPL,WuPoP,Michel2014,Turnbull,Lehmann2016,Lehmann2018}.

Although many novel potential applications based on the PDG are proposed, the physics of its formation and evolution is still not fully understood.
So far, the analytical models based on the linearization approximation of fluid equations are widely adopted in the studies of the PDG formation \cite{Sheng,Michel2013,Lehmann2019}.
In the linear fluid models, the plasma density modulation is usually assumed to be much smaller than the initial plasma density.
Under this assumption, the nonlinear terms in the fluid equations can be linearized, which leads to an  analytical solution for the plasma density modulation \cite{Sheng}.
However, the evolution of the PDG can be highly nonlinear when the plasma density peaks are extremely sharp and many times larger than the initial plasma density in the later stage.
Further, the effect of plasma temperature also plays an important role in the nonlinear dynamics and saturation of the PDG  \cite{HPeng}.
More importantly, the ion wave breaking can develop in the later stage \cite{HPeng,Forslund1979,Fri2017,Lotekar,Esarey,Zhou18}.
The ion wave breaking will bound the plasma density perturbation and lead to the final collapse of the PDG.
Until now, little attention has been paid to the PDG development at the nonlinear stage, and still less to the stage after its collapse.

In this study, we develop two theoretical models for describing the PDG  evolution beyond its linear stage or even beyond its collapse.
The first theoretical model is derived from the two-fluid plasma model using an assumption of quasi-neutrality of plasma.
This fluid model can describe the nonlinear growth process of the PDG until the sharp density peaks are as large as the initial plasma density. 
To further describe the evolution of the PDG after its collapse, the particle-mesh method  \cite{PM} is adopted in the second model.
Since the plasma is treated as individual macro-particles in the particle-mesh method, the collapse of the PDG and the PDG evolution after its collapse can be described properly.
This study extends the understanding of the whole process of the PDG evolution including its growth, saturation and collapse, which could be of great benefit to the design and analysis of related experiments.

The manuscript is organized as follows:   the modified fluid model and the particle-mesh model are developed in Sec. II. The simulation results of these two models are compared with particle-in-cell simulation results in Sec. III, with an emphasis on the saturation and collapse of the PDG in the later stage. The dependence  of the PDG saturation time on the laser intensity and the plasma density is also clarified. Finally, some discussions and a short summary are presented in Sec. IV.


\section{Models of the plasma density grating evolution}

\subsection{Fluid models}
In principle, PDGs could be induced by intersecting laser beams in plasmas in a variety of scenarios.
For simplicity, in this work the PDG is assumed to be induced by two oppositely propagating laser beams through a homogeneous plasma. One beam is propagating in the positive $x$-axis and another in the negative $x$-axis.
The laser beams are assumed to have the vector potentials $\bm{A_1}=A_1\cos(\omega_1t-k_1x)\bm{e}_y$ and $\bm{A_2}=A_2\cos(\omega_2t+k_2x)\bm{e}_y$, with the same frequency and wave number, i.e., $\omega_1=\omega_2=\omega_0$ and $k_1=k_2=k$. Here $A_1$ and $A_2$ are the electric field amplitudes of laser beams 1 and 2, respectively.
The wave number in  plasma is determined by $k=k_0\sqrt{1-{n_0}/{n_c}}$, where $k_0$ is the wave number in vacuum, $n_0$ is the background plasma density and $n_c=\omega_0^2\varepsilon_0 {m_e}/{e^2}$ is the critical plasma density corresponding to the laser frequency $\omega_0$. Here $m_e$ is the electron mass and $\varepsilon_0$ is the permittivity of free space.
The superposition of these two laser beams can form a standing wave, which will induce a ponderomotive force on the electrons.
Introducing the normalized vector potential $a_{1,2}=eA_{1,2}/m_ec^2$, this ponderomotive force can be written as \cite{Sheng}
\begin{equation}  \label{eqPonder}
\bm{F}_p={m_ec^2} a_1a_2 k \sin(2kx) \bm{e}_x.
\end{equation}
The normalized vector potential $a$ is related to the laser intensity $I$ as $a \simeq ( I  \lambda^2  \textrm{/} 1.37 \times 10^{18} [ \textrm{W} \mu\textrm{m}^2 \textrm{/cm}^2 ]  )^{1/2}$, where $\lambda$ is the laser wavelength in a vacuum.
The above equation indicates that the ponderomotive force induced by two counter-propagating laser beams has a spatial period of ${\pi}/{k}$, which will result in a spatially periodic modulation of the plasma density, i.e. the PDG formation.
In a typical linear fluid model, the density perturbation can be expressed as \cite{Sheng}
\begin{equation}
\delta n=\frac{k^2 c^2}{\omega_{p}^{2}} \frac{m_e}{m_i} a_1a_2\cos(2kx) [4\sin^2(\frac{\omega_p t}{2})-\omega_p^2 t^2], \label{EqLine}
\end{equation}
where $\omega_p=\sqrt{n_e e^2/{m_e\varepsilon_0}}$ is the plasma frequency and $m_i$ is the ion mass.
From Eq. (\ref{EqLine}), it can be found that the ion density perturbation of the PDG is always a cosine function of the $x$-coordinate.
Therefore, such a linear fluid model is applicable only for the early stage of the PDG evolution (about dozens of laser cycles) when the density  perturbation is not sharp and much smaller than the initial density.

To extend the fluid model to the nonlinear stage of the PDG evolution, one has to abandon the weak density perturbation assumption \cite{Sheng}.
Using the assumption of the plasma quasi-neutrality, we obtain a modified fluid model for describing the PDG formation, which is valid until the collapse of the PDG.
We start from the momentum equations for electrons and ions in a cold plasma
\begin{eqnarray}
n_e m_e\frac{\partial{u_e}}{\partial{t}}&=&  n_e e\frac{\partial{\varphi}}{\partial{x}}-n_eF_p,  \label{eqMomEle} \\
n_i m_i\frac{\partial{u_i}}{\partial{t}} &=& -n_i Z_i e\frac{\partial{\varphi}}{\partial{x}},  \label{eqMomIon}
\end{eqnarray}
where the convective terms are omitted for simplifying the following derivation,  $\varphi$ is the scalar potential of the space-charge field, $Z_i$ is the ion charge number, $u_i$ and $u_e$ are the fluid velocities of ions and electrons, respectively.
Since the directions of velocities and forces are all along the $x$-axis, we ignore the vector symbols of those vectors in Eqs. (\ref{eqMomEle})-(\ref{eqMomIon}) and the following derivations.

Assuming the plasma remains quasi-neutral (i.e. $n_e \equiv Z_i n_i$) in the whole process of the PDG development, the sum of the momentum equations for electrons and ions yields
\begin{equation}
m_i\frac{\partial u_i}{\partial t}=-Z_im_ec^2a_1a_2 k \sin(2 k x),  \label{eqMi} \\
\end{equation}
where the term of electron inertia is omitted since $ m_e \ll m_i$.
Normalizing the time, frequency, distance, wave number and velocity to ${2\pi}/{\omega_0}$, $\omega_0$, $\lambda$, ${2\pi}/{\lambda}$ and $c$, respectively, the above equation can be rewritten as
\begin{equation}
\frac{\partial{u_i}}{\partial t}=b\sin(hx)
\end{equation}
where $b=-2\pi a_1a_2k {Z_im_e}/{m_i} $ and $h=4\pi \sqrt{1-{n_e}/{n_c}}$.
The time integration of the above equation gives the fluid velocity for ions $u_i=b\sin(hx)t$. Substituting this into the continuity equation for ions, one can obtain the following initial-value problem
\begin{eqnarray}
\frac{\partial{n}}{\partial t}+[b\sin(hx)t]\frac{\partial{n}}{\partial x} &=& -bh\cos(hx)tn,  \label{eqIV0} \\
n(x,0)&=&n_0,
\end{eqnarray}
where $n$ is the ion density normalized to $n_c$, and $n_0$ is the initial ion density of the homogeneous plasma.
Equation (\ref{eqIV0}) is a first order quasi-linear partial differential equation, and it can be rewritten as follows
\begin{eqnarray}\label{eq}
\frac{dt}{d\tau} &=& 1,  \\
\frac{dx}{d\tau} &=& b\sin(hx)t,  \\
\frac{dn}{d\tau} &=& -bh\cos(hx)t n ,
\end{eqnarray}
where $\tau$ is an intermediate variable, and the initial condition at $\tau=0$ are $t=0$, $x=\xi$, and $n=n_0$, respectively.
By the integration of the above equations, the variables of $t$, $x$ and $n$ can be expressed as the implicit functions of $\tau$ and $\xi$ as
\begin{eqnarray}
\tau &=&  t   \label{eqIVt} \\
\ln  \left| \frac{[\tan(hx)-\sin(hx)] \tan(h \xi)\sin(h \xi) }{\tan(hx)\sin(hx)[\tan(h \xi)-\sin(h \xi)]} \right|  &=& \frac{bh}{2}{\tau}^2,  \label{eqIVx} \\
n_0\exp \left\{{\int}_0^{\tau}[-bh\cos[hx(\zeta, \xi)] \zeta ] d\zeta\right\}&=& n,  \label{eqIVn}
\end{eqnarray}
For any given time-space coordinates $(t,x)$, the corresponding intermediate variables $\tau$ and $\xi$ can be gotten from the first two equations. Substituting these two  intermediate variables into Eq. (\ref{eqIVn}), the density of ions $n$ can be obtained finally.
In the derivation of Eqs. (\ref{eqIVt})-(\ref{eqIVn}), the assumption of the weak electron density perturbation is no longer introduced.
As will be seen in the next section, therefore, this modified fluid model is capable of describing the PDG evolution in the nonlinear growth stage when the density perturbation is already as large as the initial density. 
However, this modified fluid model cannot predict the collapse of the PDG, which might be because it omits the convective terms.

\subsection{Particle-mesh model}

In order to describe the collapse of the PDG and the PDG evolution after its collapse, the second model is developed using the particle-mesh(PM) method\cite{PM}. 
In this PM model, the laser propagation is described by the wave equation, while the plasma is denoted by a large number of macro-particles that are distributed within meshes. The evolution of the PDG can be resolved by tracking the motions of these macro-particles.

As it is well known, the propagation of laser pulses in a plasma is mainly governed by the following wave equation \cite{Kruer}
\begin{equation}
(\frac{\partial ^2}{\partial t^2}-\nabla^2)a=-4\pi^2n_ea, \label{wave_equation}
\end{equation}
where the vector potential $a$, electron density $n_e$, time $t$, and space coordinate $x$ are normalized to ${m_ec^2e}$, $n_c$, ${2\pi}/{\omega_0}$, and ${2\pi c}/{\omega_0}$, respectively.  
Here, the vector potential $a$ is the stacking field of two oppositely propagating laser beams. 
In our PM simulation, these two laser pulses are loaded as the boundary condition from the left and right boundaries of the simulation box, respectively.

For tracking the macro-particle motions, let's retrospect the two-fluid model for a plasma.
For the electrons, the dimensionless continuity and motion equations can be respectively written as \cite{Kruer}
\begin{eqnarray}
\frac{\partial n_e}{\partial t}+\nabla \cdot(\bm{u}_e n_e) &=& 0, \label{electron_continue}  \\
\frac{\partial \bm{u}_e}{\partial t}+ \bm{u}_e \cdot \nabla \bm{u}_e &=& 4\pi^2\bm{E}-\frac{\nabla a^2}{2}-\frac{3v^2_{th,e} \nabla n_e}{n_e}, \label{electron_motion}
\end{eqnarray}
where $\bm{E}$ is the electrostatic field, and $\bm{u}_e$ and $v_{th,e}$ are the electron fluid and thermal velocities, respectively,
The terms in the right side of Eq. (\ref{electron_motion}) are corresponding to the electrostatic force due to the charge separation, the ponderomotive force, and the heat pressure, respectively. 
Similarly, the dimensionless continuity and motion equations for the ions can be respectively written as \cite{Kruer}
\begin{eqnarray}
\frac{\partial n_i}{\partial t}+\nabla \cdot(\bm{u}_i n_i) &=& 0, \label{ion_continue}  \\
 \frac{\partial \bm{u}_i}{\partial t}+\bm{u}_i\cdot  \nabla \bm{u}_i &=&  - \frac{4\pi^2}{m_i}\bm{E}-\frac{3v^2_{th,i}}{n_i}\nabla n_i, \label{ion_motion}
\end{eqnarray}
where $\bm{u}_i$ and $v_{th,i}$ are the ion fluid and thermal velocities, respectively. 
Since laser-induced ponderomotive force is negligible for the ions, here we only consider the electrostatic force  and the heat pressure in the ion motion equation. 

The electrostatic field $\bm{E}$ in the above motion equations is induced by the charge separation between the ions and electrons, and it is given by
\begin{equation}
\nabla\cdot \bm{E}=n_i-n_e. \label{electric_field}
\end{equation}

Although the fluid equations (\ref{electron_continue})-(\ref{ion_motion}) can be numerically solved  directly in Euler coordinates to get the time evolution of the plasma, the effects of ion wave-breaking and particle trapping will be lost in this way \cite{Zhou18,L. Yin}. As a result, the collapse of the PDG and the PDG evolution after the collapse can not be described by solving these fluid equations directly in Euler coordinates. 
Therefore, we adopt the PM method to resolve the electron and ion motions, in which the ion wave-breaking can be treated self-consistently \cite{PM}.
In the PM model, the simulation box is divided into fixed meshes. Then the electrons and ions are denoted by macro-particles that distributed within these meshes.
The macro-particle motions are governed by the total force $F$ that defined by the right terms in the motion equations  (\ref{electron_motion}) and (\ref{ion_motion}) for the electrons and ions, respectively.
In our numerical scheme, the force is firstly calculated on the meshes and then interpolated to the position of each macro-particle. 
Subsequently, the velocity and position of each macro-particle can be updated as
\begin{eqnarray}\label{particle_move}
v_i(t) &=& \int_0^t F(x_i,t')dt'+v_{i,0} ,  \\
x_i(t) &=& \int^t_0 u_i(t')dt'+x_{i,0},
\end{eqnarray}
where $F(x_i,t')$ is the total force acting on the $i$-th macro-particle at time $t'$, $v_i(t)$ and $x_i(t)$ are the velocity and position of the $i$-th macro-particle at time of $t$, and $v_{i,0}$ and $x_{i,0}$ are the initial velocity and position of the $i$-th macro-particle, respectively. 
In our numerical scheme,  the second order Runge-Kutta algorithm is adopted to update the velocities and positions of macro-particles. 

After updating the velocities and positions of macro-particles, the electron and ion densities on the meshes can be updated by the interpolation of macro-particles to each mesh as done in the PIC code. Then, the electrostatic field and force can be calculated according to Eq. (\ref{electric_field}). 

After the electron density is updated, the vector potential of the laser pulses can be updated according to the wave equation (\ref{wave_equation})  using 2-order central difference scheme for both time and space. Consequently, the ponderomotive force can be updated. 

Combining the updated electron and ion densities, electrostatic and ponderomotive forces, the total force upon each macro-particle can be updated in return for the next iteration.

Since the macro-particles in the PM model can move cross each other freely, this PM model is able to capture the effects of wave-breaking and particle trapping as typical PIC codes do.  
Further, the effects of thermal pressure on the evolution of the PDG are also included  in the motion equations  (\ref{electron_motion}) and (\ref{ion_motion}) for the electrons and ions, respectively.

The main difference between our PM model and typical PIC codes is that the force acting on the macro-particles is calculated according to the fluid motion equation in the PM model rather than the electromagnetic field in the PIC codes. 
Moreover, the wave equation is solved for the laser propagation in the PM model while the Maxwell equations are solve in the typical PIC codes.
Therefore, the PM model runs more efficiently than the typical PIC code. 
With the same macro-particle number per mesh (or cell) and the same steps in time and space, the computation efficiency of the PM model is about three times higher than that of the PIC simulation. More importantly, we find that much less macro-particles (only several particles per mesh) are required in our PM model than those (dozens to hundreds of particles per cell) in the typical PIC simulations in order to get a similar numerical precision. 
In addition, the numerical heating that usually appears in PIC codes \cite{EstelleC} seems to be not obvious in the PM model.

\section{Comparison with particle-in-cell simulations}

\begin{figure}
    \centering\includegraphics[width=0.5 \textwidth]{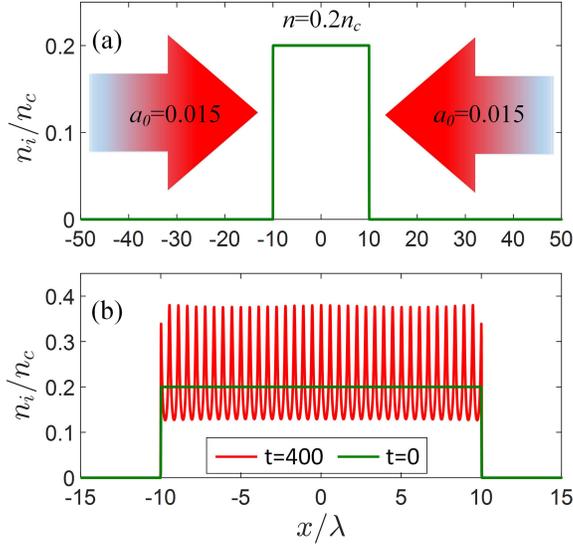}
	\caption{
		(a) Laser and plasma parameters used in the PIC simulation. The initial plasma is cold and has a uniform density $n_0=0.2n_c$ in $|x| \le 10 \lambda$. Two linearly polarized laser pulses with the same frequency $\omega_0$ and amplitude $a_0=0.015$ are launched from two boundaries, as shown by the arrows.
		(b) The ion density profile obtained from the PIC simulation at $t=400 T_0$ is compared with that at $t=0$, where $T_0=2\pi/\omega_0$ is the laser oscillation period.
	}\label{figConfig}
\end{figure}

To verify the proposed models, the evolution processes of the PDGs calculated by different theoretical models are compared with that obtained from PIC simulations.
One-dimensional PIC simulations are conducted using the code Osiris \cite{Fonseca}.
A simulation box with a dimension of  $100\lambda$ is located at $x\in[-50\lambda,50\lambda]$, and a homogeneous plasma with a density of $n=0.2n_c$ is located at the central region of $-10\lambda \le x \le 10\lambda$.
As shown in Fig. \ref{figConfig}(a), two linearly polarized laser pulses with the same frequency ${\omega_0}$ and same amplitude of $a_0=0.015$ are launched from the left and right boundaries of the simulation box.
The cell size is chosen as $0.01\lambda$ with 100 macro particles per cell.
For convenient comparison with the theoretical models, the laser pulses in the simulation have flat-top profiles and they are long enough so that the PDG has the time to develop, saturate and collapse.
The initial plasma is assumed to be cold and the ions are protons.
The same simulation box, laser and plasma parameters are used in the PM model as those in the PIC simulation. 
However, we find that 5 macro-particles per mesh is enough to obtain a good numerical precision in the PM model.

As shown in Fig. \ref{figConfig}(b), the ion density profile at $t=400 T_0$ obtained from the PIC simulation confirms the formation of the PDG. The peak density of the PDG at this moment is about twice the uniform density at $t=0$.

\begin{figure}
	\centering\includegraphics[width=0.5 \textwidth]{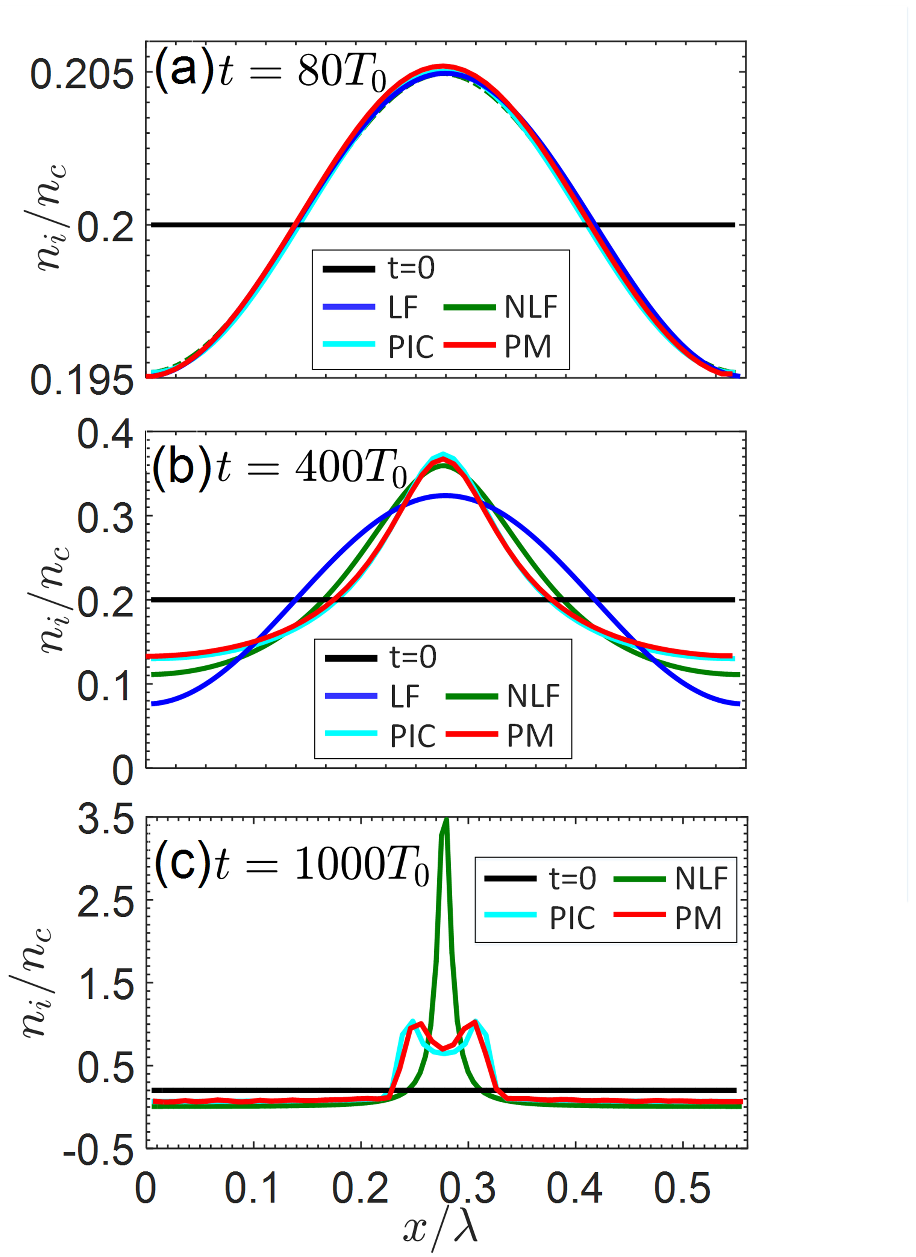}
	\caption{
		The ion density profiles of a single cycle of the PDG at the center region  $0< x < \pi/k$  of the simulation box in (a) the early linear growth stage at $t=80T_0$, (b) the nonlinear growth stage at $t=400T_0$, and (c) the stage after the wave-breaking at $t=1000T_0$, respectively.
	  The comparison is made among the results from the linear fluid model (Eq. \ref{EqLine}) \cite{Sheng} (labeled as ``LF"), nonlinear fluid model (``NLF"), particle-mesh model (``PM"), and  PIC simulation (``PIC"). The black lines indicate the ion density profile at  $t=0$.
	}\label{figProfiles}
\end{figure}

To compare the different models in detail, we zoom in a single cycle of the PDG at the center region  $0< x < \pi/k$ of the simulation box, where $\pi/k= \lambda/ 2\sqrt{1-n/n_c} \simeq 0.56 \lambda$.
In Fig. \ref{figProfiles}, the ion density profiles at this zoom region obtained from three theoretical models are compared with that from the PIC simulation.

For the early stage of the PDG evolution (at around $t=80T_0$), it can be found that the PDG density profiles calculated by the linear fluid model of Eq. (\ref{EqLine}), modified fluid model of Eq. (\ref{eqIVn}),  PM model and PIC simulation are all in a good agreement as show in Fig. \ref{figProfiles}(a). 
More importantly, the PDG density profile at this time is nearly a cosine function.

If the ion density modulation is comparable to the initial density in the later stage, however, the PDG density profile will no longer be a cosine function. Consequently, the linear fluid model is not able to accurately predict the PDG evolution as show in Fig. \ref{figProfiles}(b).
In contrast, the ion density profiles predicted by our modified fluid model and PM model are still in good agreement with the PIC simulation result in Fig. \ref{figProfiles}(b).

More importantly, the PIC simulation shows that the peaks of the PDG will split with the increasing of the peak density as shown in Fig. \ref{figProfiles}(c).
This highlights that the PDG will saturate, and its periodic structure will be finally destroyed due to ion wave-breaking.
However, Fig. \ref{figProfiles}(c) shows that the PDG peak density calculated by our modified fluid model using Euler coordinates would increase continuously.
In contrast, the ion wave breaking is well captured by the particle-mesh model in which the plasma is described as a collective of individual macro-particles.
As a result, the collapse of the PDG is reproduced by the particle-mesh model as shown in Fig. \ref{figProfiles}(c).

\begin{figure}
	\centering\includegraphics[width=0.5 \textwidth]{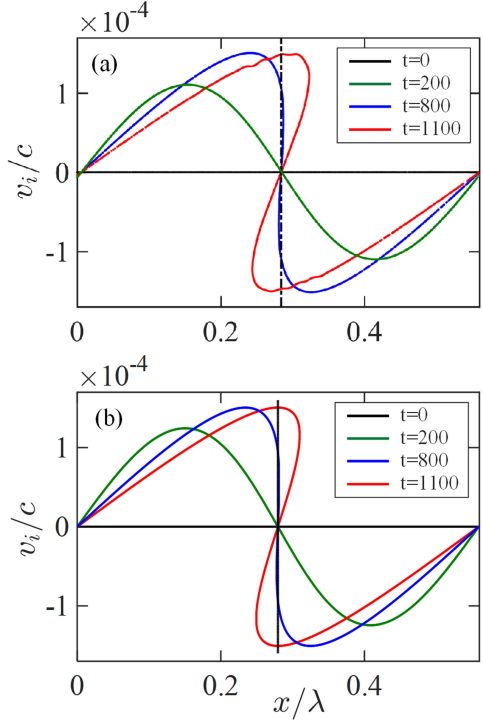}
	\caption{
		The ion distributions in the $x-v_x$ phase space obtained from (a)  PIC simulation and (b) PM model at some typical times: $t=0$ as the initial state, $t=200T_0$ in the growth stage,  $t=800T_0$ around the saturation time when ion wave-breaking occurs, and $t=1100T_0$ after the PDG collapse.
	}\label{figPhase}
\end{figure}

To better understand the saturation and collapse of the PDG, the ion distributions in the $x-v_x$ phase space obtained from the PIC simulation and PM model at some typical times are displayed in Fig. \ref{figPhase}.
Under the periodic ponderomotive force of two oppositely propagating laser beams, the velocities of ions in the left half region for the chosen cycle of PDG is positive while the velocities of ions are negative in the right half region, as shown in Fig. \ref{figPhase}.
Therefore, the density at the center of a cycle of PDG increases and the PDG develops.
With the increasing of the peak density, the slope of the ion phase-space distribution at the density peak goes to negative infinity and then reverses its sign at around $t\simeq 800 T_0$,  where ion trajectories begin to cross each other.
Due to the inertia effect, the ions from the left and right half parts will continue to move across each other.
As a result, the ion fluid velocity at a given position, such as the position of the density peak, becomes no longer unique after $t\simeq 800 T_0$.
In other words, the wave breaking takes place and saturates the PDG.
The similar ion distributions in the $x-v_x$ phase space were previously reported as a character of the wave breaking \cite{Forslund1979,J. G. Wang}. 
Since our PM model can capture the ion wave breaking, the ion phase space distribution obtained from the PM model agrees well with that from the PIC simulation at the whole process of the PDG evolution as shown in Fig. \ref{figPhase}(a) and Fig. \ref{figPhase}(b).

\begin{figure}
	\centering\includegraphics[width=0.5 \textwidth]{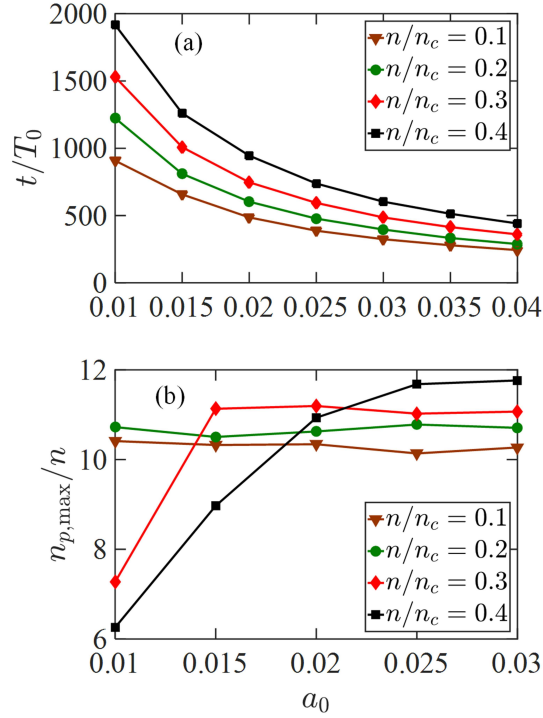}
	\caption{
		(a) The saturation time and (b) the maximal achievable peak ion density $n_{p,\max}$ of the PDG obtained from the PM model with different laser intensities $a_0$ and initial plasma densities $n$. Except for the laser intensities and plasma densities, other laser-plasma parameters are the same as those used in Figs. \ref{figProfiles}.
	}\label{figScaling} \label{Peak_density}
\end{figure}

To study the dependence of the growth rate of the PDG on the plasma density and the laser intensity, we calculate the saturation time of the PDG by the particle-mesh model under different plasma densities and laser intensities.
The results are displayed in Fig. \ref{figScaling}(a). Except for the laser intensities and plasma densities, other laser-plasma parameters are the same as those used in Fig. \ref{figProfiles}.
From Fig. \ref{figScaling}(a), it can be found that the saturation time $T_s$ of the PDG decreases gradually with an increasing laser intensity $a_0$ for a given plasma density $n_0$.
While the saturation time $T_s$ of the PDG increases with an increasing plasma density $n_0$ for a given laser intensity.
This is because the saturation will be achieved faster with a stronger ponderomotive force, and Eq. (\ref{eqPonder}) indicates that the ponderomotive force increases with increase in the laser intensity $a$ and the wave number $k$ in plasma, while $k$ decreases with increase in the plasma density $n_0$.
It is worth pointing out that the saturation time of the PDG also depends on other parameters such as the plasma temperature, the ion mass ${m_i}$ and so on.
From a large number of calculations based on the particle-mesh model, the saturation time $T_s$ of the PDG for a cold plasma can be roughly fitted by
\begin{equation}
T_s=\left( 0.73 + \frac{M}{3.7} \right) (21.09n+6.97)a^{-0.16n-0.98},
\end{equation}
where $M={m_i}/{m_p}$ is the ion mass normalized to the mass of the proton, and $a$ and $n$ are the normalized laser intensity and initial plasma density, respectively.
The above equation can be conveniently used to evaluate the saturation time of the PDG in experiments.

As soon as the PDG begins to collapse, its peak density will reach a maximum value.
In Fig. \ref{Peak_density}(b), the ratio of this maximum achievable peak density $n_{p,\max}$ to the initial plasma density $n$ is shown as a function of the laser intensity and initial plasma density. 
It can be found that the maximum achievable peak density tends to become saturated with the increasing of the laser intensity. 
More interestingly, the saturation value of the ratio $n_{p,\max}/n$ ($\sim 10$) seems not sensitive to the initial plasma density.

\section{Discussion and Conclusion}

\begin{figure}
	\centering\includegraphics[width=0.5 \textwidth]{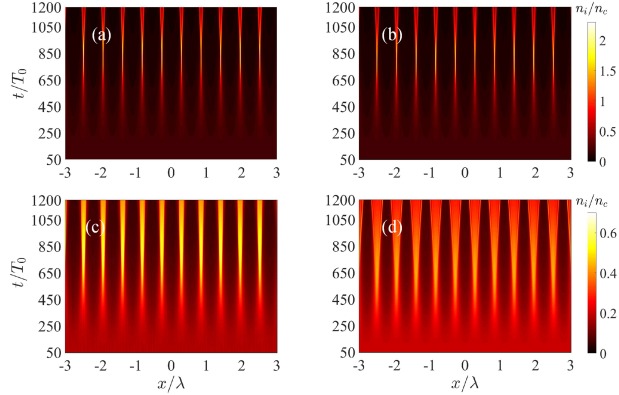}
	\caption{
		The time evolution of the PDG at the region $|x| \le 3\lambda$ obtained from PIC simulations (left column) and PM model (right column), respectively. Both electron and ion temperatures are set to be zero in the upper panels (a) and (b), while  $T_e=10$ eV and $T_i=1$ eV are used in the lower panels (c) and (d).  Except for the plasma temperature, other  laser-plasma parameters are the same as those used in Fig. (\ref{figProfiles}).
	}\label{time_evolution}
\end{figure}

Since the thermal effect could play an important role in the PDG evolution, we compare the PDG evolution in two plasmas with zero and nonzero temperatures, respectively. 
As shown in Fig. (\ref{time_evolution}), it is verified both by our PM model and PIC simulations that the peak density of the PDG will decease due to the heat pressure in a plasma with nonzero temperature. Meanwhile, the PDG evolution becomes faster and collapses earlier with a nonzero plasma temperature. 
If the plasma temperature is sufficiently high, the PDG formation can even be prevented by the thermal pressure \cite{HPeng}. 
Therefore, to obtain a quasi-steady PDG, the plasma temperature should be controlled to be not too high.

It is worth pointing out that neither our PM model nor the employed PIC code considers the collisional absorption. If the collisional absorption is taken into account self-consistently, the PDG evolution is expected to become faster since the plasma will be heated gradually. In the future work, we will update our PM model to include the collisional absorption self-consistently, which would play an important role in the more accurate estimation of laser and plasma parameters for the formation of a stable PDG.

\begin{figure}
	\centering\includegraphics[width=0.5 \textwidth]{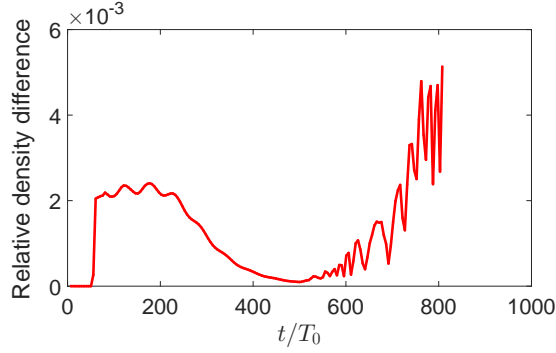}
	\caption{
		The time evolution of the maximal relative density difference between the electrons and ions obtained from the PIC simulation.
	}\label{figNdiff}
\end{figure}

A basic assumption adopted in our modified fluid model is that the plasma remains quasi-neutral for the entire process of the PDG formation.
To verify this assumption, the density difference between the electrons and ions is monitored in the PIC simulation.
Defining the maximal relative density difference as
\begin{equation}
\Delta n_{\max}=\max \limits_{|x| \le 10} \left| \frac{n_e-n_i}{n_i} \right|,
\end{equation}
the time evolution of this difference obtained from the PIC simulation is show in Fig. \ref{figNdiff}.
It is confirmed that the maximal relative density difference between the electrons and ions is always less than $1\%$ for the entire process of the PDG formation, which shows the assumption of  ``quasi-neutral'' is reasonable.

In summary, the time evolution of the PDG induced by intersecting laser beams is studied by two newly-constructed nonlinear theoretical models.
The first is a nonlinear fluid model, in which a set of first order quasi-linear partial differential equations is derived from the fluid equations without the linearization approximation.  This set of first order partial differential equations can be used to predict the time evolution of the PDG beyond the linear growth stage, but still before the ion wave breaking.
In the second model, the particle-mesh method is adopted to describe the ion wave-breaking of PDG.
Considering the wave-breaking effect, it is found that the peak density of the PDG will decrease after it reaches a maximum value. 
Since the wave breaking is well treated using this particle-mesh model, it can describe the time evolution of the PDG beyond the saturation time.
Further, the dependence of the saturation time of the PDG on the laser intensities $a_0$ and plasma densities $n_0$ is investigated using this particle-mesh model.
It is found that the saturation time of the PDG increases with the plasma density and decreases with the laser intensity.
Our study indicates that it is possible to produce the PDG with a life time on the order of picoseconds, which can be used to manipulate intense laser pulses with duration ranging from picoseconds down to femtoseconds.

\begin{acknowledgments}
The work was supported by the National Natural Science Foundation of China (Grant Nos. 11975154, 11675108, 11655002, 11721091, 11535001, and 11775144), Presidential Foundation of the Chinese Academy of
Engineering Physics (No. YZJJLX2016008),  the Strategic Priority Research Program of Chinese Academy of Sciences (Grant No. XDA25050100),  Science Challenge Project (No.TZ2018005) and EPSRC (Grant No. EP/R006202/1).
Simulations have been carried out on the Pi supercomputer at Shanghai Jiao Tong University.
\end{acknowledgments}

\section*{Data Availability Statement}
The data that support the findings of this study are available from the corresponding author upon reasonable request.


\end{document}